# A framework for Iot-Enabled Smart Agriculture

**Authors**: Nsengiyumva Wilberforce and Dr. Mwebaze Johnson

Makerere University

**Abstract**

Unpredictable weather patterns and a lack of timely, accurate information significantly challenge farmers in Uganda, leading to poor crop management, reduced yields, and heightened vulnerability to environmental stress. This research presents a framework for IoT-enabled smart agriculture, leveraging Raspberry Pi-based technology to provide real-time monitoring of weather and environmental conditions. The framework integrates sensors for temperature, rainfall, soil moisture, and pressure, connected via an MCP3208 analog-to-digital converter. Data is displayed on an LCD for immediate feedback and transmitted to the ThingSpeak platform for centralized storage, analysis, and remote access through a mobile app or web interface. Farmers can leverage this framework to optimize irrigation schedules and improve crop productivity through actionable insights derived from real-time and forecasted data on rainfall, temperature, pressure and soil moisture. Additionally, the system incorporates predictive weather forecasting to dynamically control sensor activity, reducing energy consumption and extending sensor lifespan. Simulated using Proteus, the proposed framework demonstrates significant potential to mitigate the impacts of unpredictable weather by reducing water consumption, improving forecasting accuracy, and boosting productivity.

## 1. INTRODUCTION

Weather and environmental monitoring systems are indispensable for enhancing agricultural productivity and mitigating climate-related risks worldwide. Advanced systems such as Uganda's Weather Information Dissemination System (WiDS) [1], AirQo [2], which focuses on air quality monitoring in African urban centers, and globally recognized platforms like NOAA's National Weather Service in the United States [3], the UK Met Office [4], and the India Meteorological Department (IMD) [5] have been instrumental in providing critical weather forecasts and environmental data. These systems play a crucial role across sectors such as urban planning, public health, disaster management, and aviation. However, their application in addressing the specific needs of the agricultural sector remains limited [6].

Farmers worldwide require focused and actionable insights into environmental parameters that directly impact their agricultural practices. Key parameters include soil moisture, crucial for determining irrigation schedules and ensuring optimal crop health [7]; rainfall, vital for planning planting schedules, pest control strategies, and water resource management; and temperature and atmospheric pressure, essential for predicting crop stress, identifying weather anomalies, and tailoring farming practices to local conditions. Despite the critical importance of these parameters, most existing systems prioritize generalized services over actionable agricultural insights [8], [9].

This challenge is particularly pronounced in low-income and rural regions, where farmers often lack access to cost-effective tools for real-time environmental monitoring and tailored data visualization. Efforts to address these gaps include platforms such as MeteoSwiss in Switzerland, JAXA's Himawari-8 satellite system in Japan, and the European Centre for Medium-Range Weather Forecasts (ECMWF), along with initiatives like Weather Underground, OpenWeatherMap, and agriculture-focused IoT projects such as John Deere's Precision Agriculture Tools. While these systems provide valuable weather and climate data, few are specifically designed to meet the granular needs of farmers.

This research proposes **a framework for IoT-enabled smart agriculture**, designed to bridge these gaps and provide farmer-centric solutions leveraging Internet of Things (IoT) technology. The framework utilizes platforms like ThingSpeak for real-time environmental monitoring and data visualization [10], focusing on key agricultural parameters such as soil moisture, rainfall, temperature, and atmospheric pressure. These insights empower farmers to optimize planting and harvesting schedules, improve irrigation management by minimizing water wastage, and enhance crop yields. Additionally, the framework addresses risk mitigation for unpredictable weather by enabling farmers to respond proactively to extreme events [12].

The paper is organized as follows: Section 2 reviews related research works and highlights how this framework addresses gaps in the current state of the art. Section 3 details the proposed system architecture, which combines sensor arrays, communication technologies, and applications to provide automated decision-making support. Section 4 describes how the framework was experimentally validated and tested through simulation. Section 5 presents simulation results under various conditions, and Section 6 concludes the research findings.

## 2. RELATED WORKS

The Internet of Things (IoT) has emerged as a cornerstone in transforming traditional agricultural practices into precision agriculture by enabling real-time monitoring and control of farming activities[13]. IoT devices such as soil sensors, weather stations, and smart irrigation systems have been extensively researched and proven to optimize resource utilization and improve crop yields[14]. For example, research by Suprava et al. demonstrates how IoT-based soil sensors can accurately measure moisture levels, enabling automated irrigation systems to reduce water wastage significantly[15]. Additionally, IoT weather stations collect critical environmental data, such as temperature and humidity, allowing farmers to make informed decisions about planting and harvesting schedules. Smart agriculture platforms also employ IoT-enabled drones to monitor crop health and detect pest infestations, a method validated by Arnab et al. to significantly increase efficient crop management due to systematic monitoring.

Recent studies have also highlighted the integration of IoT with cloud computing and big data analytics to enhance the scalability and efficiency of precision agriculture. Research by Lee et al. explores the role of IoT in collecting large-scale agricultural data and utilizing cloud-based platforms for processing and predictive analytics, enabling farmers to anticipate crop diseases and forecast yields with higher accuracy. In Africa, Amsale et al.[16] Emphasize the impact of IoT on smallholder farmers, where IoT and WSN prove to be assisting tool in addressing information needs, improving decision-making and market access. Despite these advancements, challenges such as high initial costs, limited network coverage in rural areas, and data security concerns persist, as highlighted by Ahmed et al., suggesting a need for policy interventions and collaborative efforts to bridge the gap in IoT adoption in agriculture.

## 3. SYSTEM ARCHITECTURE

The smart agriculture system architecture (Figure 3-1) comprises several interconnected layers that collaboratively collect, process, and disseminate weather-related data to farmers. At the foundation of this system is the sensor layer, which includes various sensors designed to monitor environmental variables such as temperature, pressure, soil moisture, and rain. These sensors are connected to the MCP3208 ADC, which converts the analog sensor data into a digital format readable by the Raspberry Pi. This layer plays a crucial role in gathering raw environmental data essential for further processing.

**Figure 3-1** presents a detailed illustration of the architecture, showcasing the interaction and integration of the system's components.

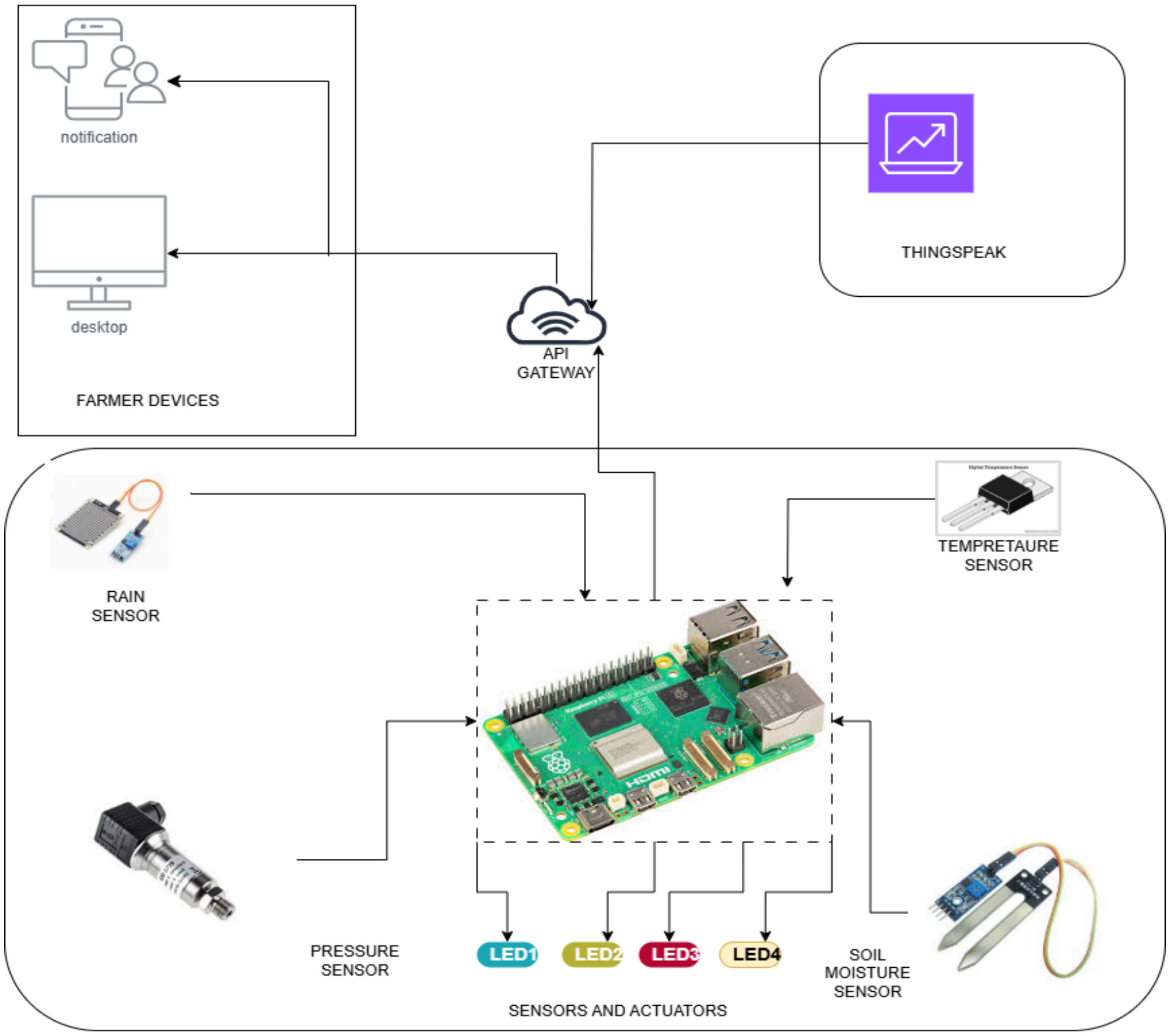

*Figure 3-1SIMULATION ARCHITECTURE*

The processing layer serves as the central hub of the system, with the Raspberry Pi acting as its core. This layer collects data from the sensors, processes it, and transmits it to a cloud platform called ThingSpeak for further analysis. Communication between the Raspberry Pi and the sensors is facilitated through GPIO (General Purpose Input/Output) pins, ensuring a continuous flow of data within the system. By managing the initial processing and data flow, this layer ensures smooth integration between the physical and digital components of the architecture.

Once the data is processed, it is transmitted to the cloud and data analysis layer. This layer relies on ThingSpeak, a cloud-based platform that stores and analyzes the collected data. ThingSpeak provides real-time analytics and visualizations of weather data, enabling farmers to monitor trends and receive timely weather alerts.

The application layer provides farmers with direct access to the processed data. Through the ThingSpeak platform, users can view their real-time data in the form of graphs and other formats, ensuring that the information is both actionable and easily understandable. In addition to web-based access, this layer also incorporates a mobile phone application that delivers alerts directly to users' devices. These mobile notifications provide real-time weather updates, ensuring that farmers receive critical information promptly, even when they are away from their computers. This integration of mobile alerts enhances the accessibility and usability of the system, allowing farmers to make informed decisions anytime and anywhere. Together, these layers form an integrated IoT architecture that empowers farmers with reliable, timely, and easily accessible weather information to support agricultural decision-making.

### 3.1. TOOLS AND TECHNOLOGIES

The IoT system architecture (Figure 2-1) described above relies on various tools and technologies to function seamlessly and deliver accurate, real-time weather information to farmers. These tools and technologies are carefully chosen to ensure the system's reliability, scalability, and ease of use. At the hardware level, the system employs sensors such as temperature, pressure, soil moisture, and rain sensors, which are vital for monitoring environmental variables. These sensors interface with the MCP3208 ADC, a component responsible for converting analog data into digital signals that can be processed by the Raspberry Pi. The Raspberry Pi itself serves as the system's processing hub, enabling the integration of sensor data and ensuring its smooth transmission to the cloud. Additionally, the GPIO (General Purpose Input/Output) pins of the Raspberry Pi facilitate communication with the sensors, maintaining a consistent data flow.

On the software side, the system leverages ThingSpeak, a powerful cloud-based platform designed for IoT applications. ThingSpeak is used to store, analyze, and visualize the weather data collected by the sensors. With its real-time analytics and visualization capabilities, ThingSpeak enables farmers to monitor trends and receive actionable insights through an intuitive interface. Moreover, the platform supports remote monitoring, making it a crucial component of the system's cloud and data analysis layer.

To enhance accessibility, the system includes a mobile phone application that ensures farmers receive weather alerts directly on their devices. The mobile app is designed to complement the web-based interface provided by ThingSpeak, offering users a flexible and convenient way to access critical weather data. This integration of mobile alerts ensures that farmers are always informed, regardless of their location.

By combining robust hardware components with advanced software tools, the IoT system architecture ensures accurate data collection, efficient processing, and effective dissemination of weather-related information. These technologies work in harmony to empower farmers with the tools they need to make informed decisions, improving agricultural productivity and resilience.

## 4. MODEL DEVELOPMENT/SIMULATION

The simulation depicted in Figure 4-1 provides an in-depth representation of the interaction and functioning of key components within the IoT system architecture. This simulation highlights how data is captured, processed, transmitted, and analyzed to deliver actionable insights for users. The various components such as the gateway, display module, SPI device, sensors, actuators, and the ThingSpeak platform work together to ensure seamless data flow and system responsiveness.

**Figure 4-1** illustrates the architecture of this simulation, detailing the roles of each component as described below:

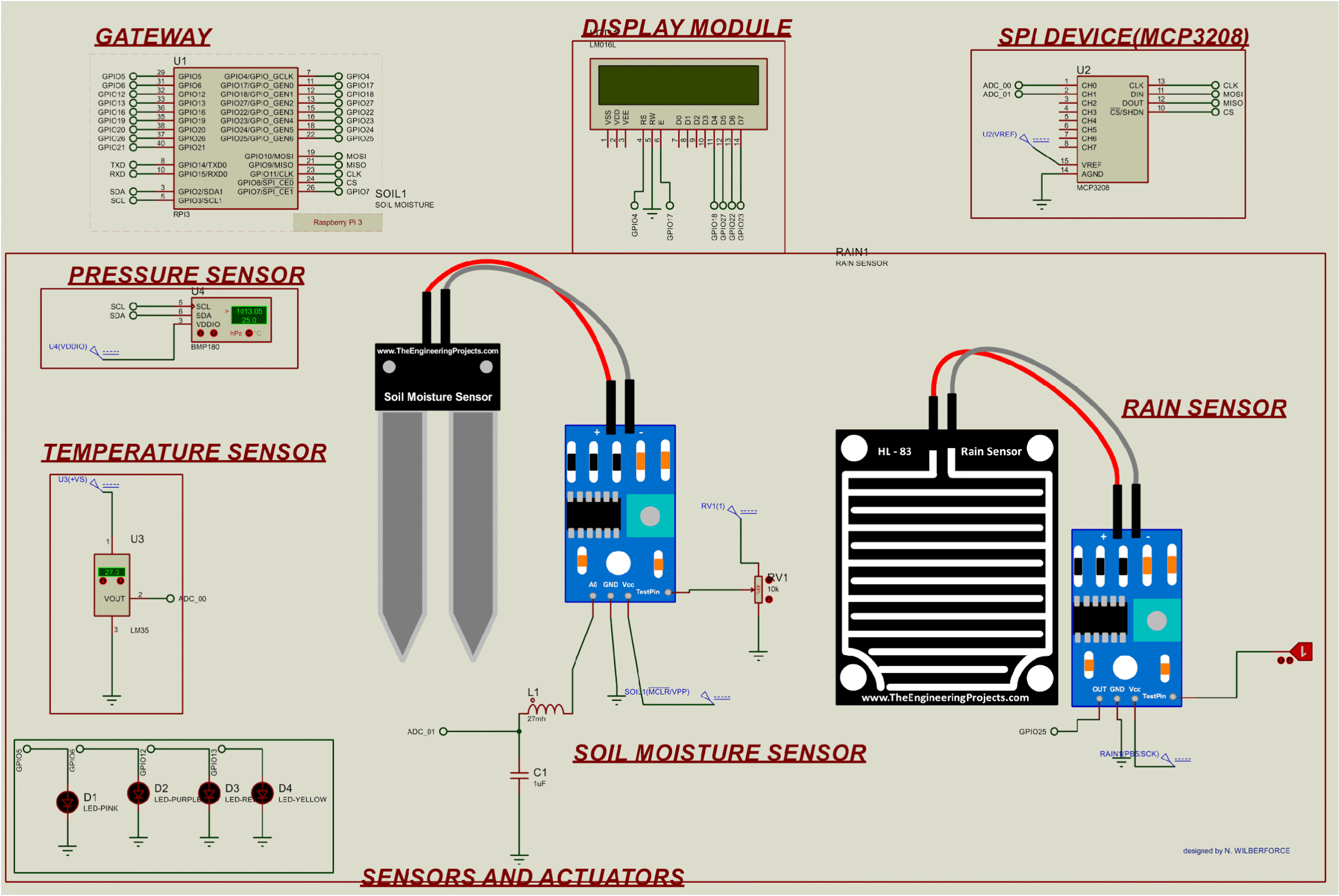

*Figure 4-1-SMART AGRICULTURE SIMULATION*

The **gateway** serves as a critical intermediary between the sensor layer and the cloud platform. It consists of a Raspberry Pi microcontroller, which receives data captured by the sensors, processes it, and transmits it to the cloud using its inbuilt Wi-Fi module. This Wi-Fi connectivity ensures seamless and reliable communication with the ThingSpeak platform, enabling real-time data transfer. For this simulation, the captured data is sent to ThingSpeak for storage, analysis, and the generation of meaningful insights that can assist users in decision-making.

The **display module** plays a vital role in presenting the data collected by the sensors. It uses SPI (Serial Peripheral Interface) to transfer sensor readings to the Raspberry Pi, allowing the data to be displayed on an LCD screen. This ensures local, real-time visibility of environmental parameters, which is valuable for on-site monitoring and immediate validation of the system's functionality.

The **SPI device**, specifically the MCP3208, acts as the communication medium that bridges the sensors and the gateway. This device converts the analog data received from the sensors into digital signals, enabling the Raspberry Pi to process and interpret the information accurately. With its eight input channels, the MCP3208 can interface with multiple sensors simultaneously, making it ideal for applications that require monitoring of diverse environmental parameters such as temperature, pressure, soil moisture, and rainfall. This capability aligns seamlessly with the multi-layered architecture described earlier.

The system also incorporates **sensors and actuators**, which are essential for capturing environmental data and responding to changing conditions. The sensors include a temperature sensor, pressure sensor, soil moisture sensor, and rain detector. Actuators, such as LED lights, are used to indicate specific environmental states by turning on or off based on predefined conditions. This dynamic interaction ensures that the system not only monitors but also responds to environmental changes effectively.

The **ThingSpeak platform** functions as the cloud-based hub for this IoT system. It facilitates the storage, analysis, and visualization of the data received from the gateway. With its robust analytics and real-time alerting capabilities, ThingSpeak provides users with actionable insights. Additionally, the platform allows for remote monitoring, ensuring that users can access the system's outputs anytime and anywhere via mobile or web-based interfaces.

In addition to the above, the Raspberry Pi's **inbuilt Wi-Fi module** ensures that the system remains connected to the cloud platform at all times. This connectivity is critical for enabling remote access and ensuring that the analyzed data is updated in real time. Furthermore, the architecture takes into account the need for **power management**, as sensors and actuators require a stable power supply to function effectively. The scalability of the system also ensures that more sensors or actuators can be added to meet future requirements, making it adaptable for larger-scale deployments.

## 5. SIMULATION RESULTS AND ANALYSIS

This section presents the results obtained from the IoT system simulation, which models environmental parameters such as **temperature**, **pressure**, **soil moisture**, and **rainfall detection**. The simulation aimed to demonstrate how the system collects, processes, and communicates these weather conditions, providing actionable insights for farmers through the cloud and mobile app. The following results illustrate how the system behaves under different simulated weather conditions, including pressure fluctuations and rainfall events.

### 5.1. SIMULATION SETUP AND PARAMETERS

For the simulation, several environmental parameters were considered to replicate typical weather conditions experienced by farmers. The temperature was simulated to range between 20°C and 40°C, reflecting daily and seasonal cycles. Atmospheric pressure values were set between 1010 hPa and 1025 hPa to account for fluctuating weather patterns. Soil moisture levels were simulated between 10% and 50%, responding to rainfall events and evapotranspiration. Rainfall detection was modeled to detect presence of rain or no rain, enabling the system to detect rainfall and trigger relevant actions. These parameters served as inputs for the system's processing unit, the Raspberry Pi, which relays the processed data to the cloud platform for further analysis and visualization.

### 5.2 SIMULATED DATA AND RESULTS

#### 5.2.1 TEMPERATURE DATA

Figure 5-1 shows the temperature variations over a 1 hour simulation period.

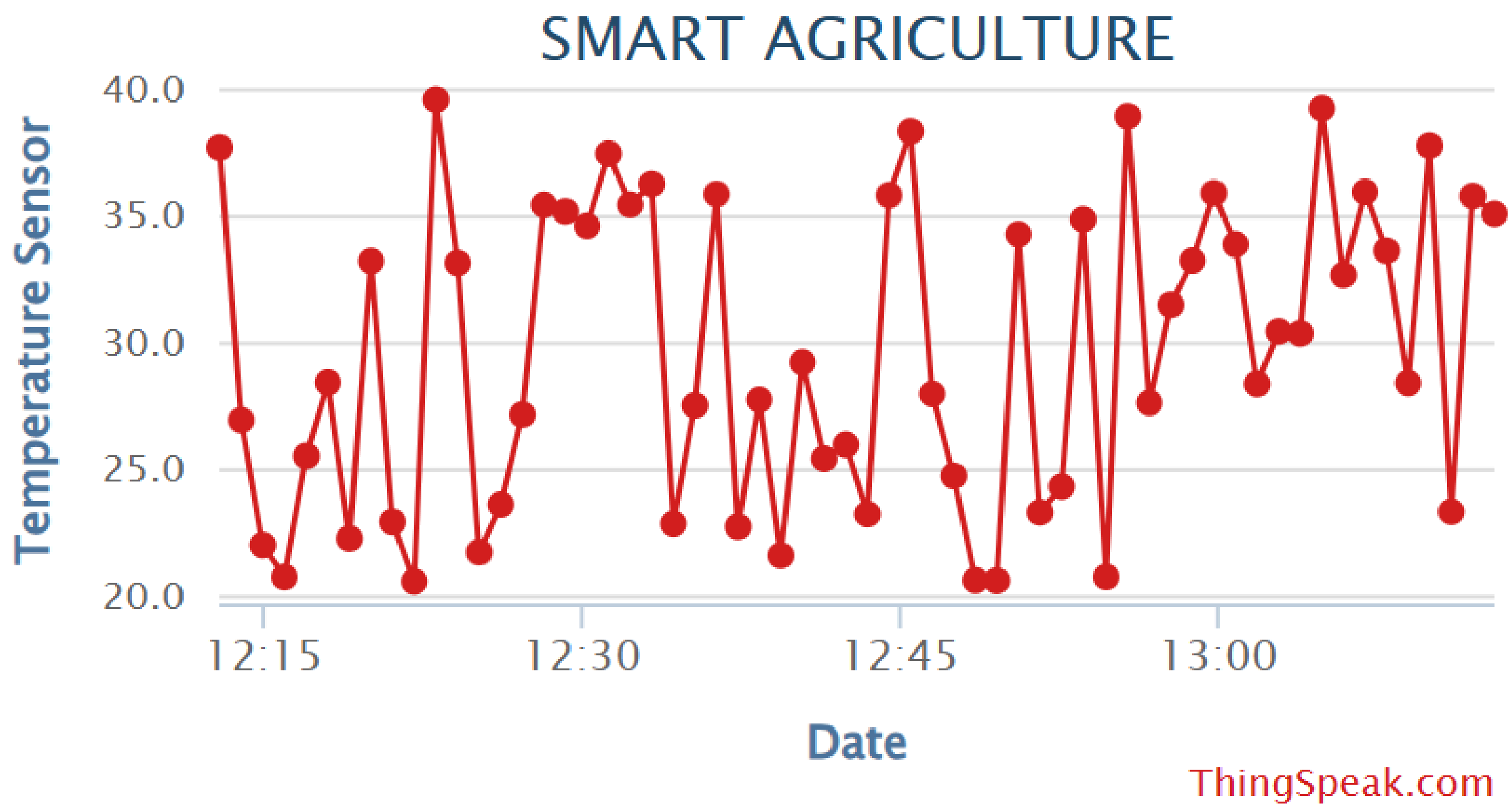

*Figure 5-1Simulated Temperature Variation Over 1 Hour*

In **Figure 5-1**, the graph illustrates the temperature variations over time, as simulated from the Proteus environment. The data highlights distinct environmental phases. Initially, between the first and third minutes, the temperature rises sharply, peaking at approximately 37.72°C. This period could simulate a drought scenario, with high temperatures potentially indicating low moisture levels and increased crop stress. Subsequently, between the fourth and fifth minutes, the temperature drops significantly to around 22.04°C, simulating a rain event or the onset of cooling conditions that could alleviate drought stress and provide favorable conditions for plant growth. These fluctuations, as captured in the graph, emphasize the importance of real-time environmental monitoring to identify critical weather patterns and their impact on agricultural practices. Such insights can guide decisions on irrigation scheduling and other farming activities, aligning with the research's goal of developing IoT-based systems for actionable agricultural insights.

#### 5.2.2 PRESSURE DATA

Figure 4-2 displays the atmospheric pressure variations during the same 1 hour period.

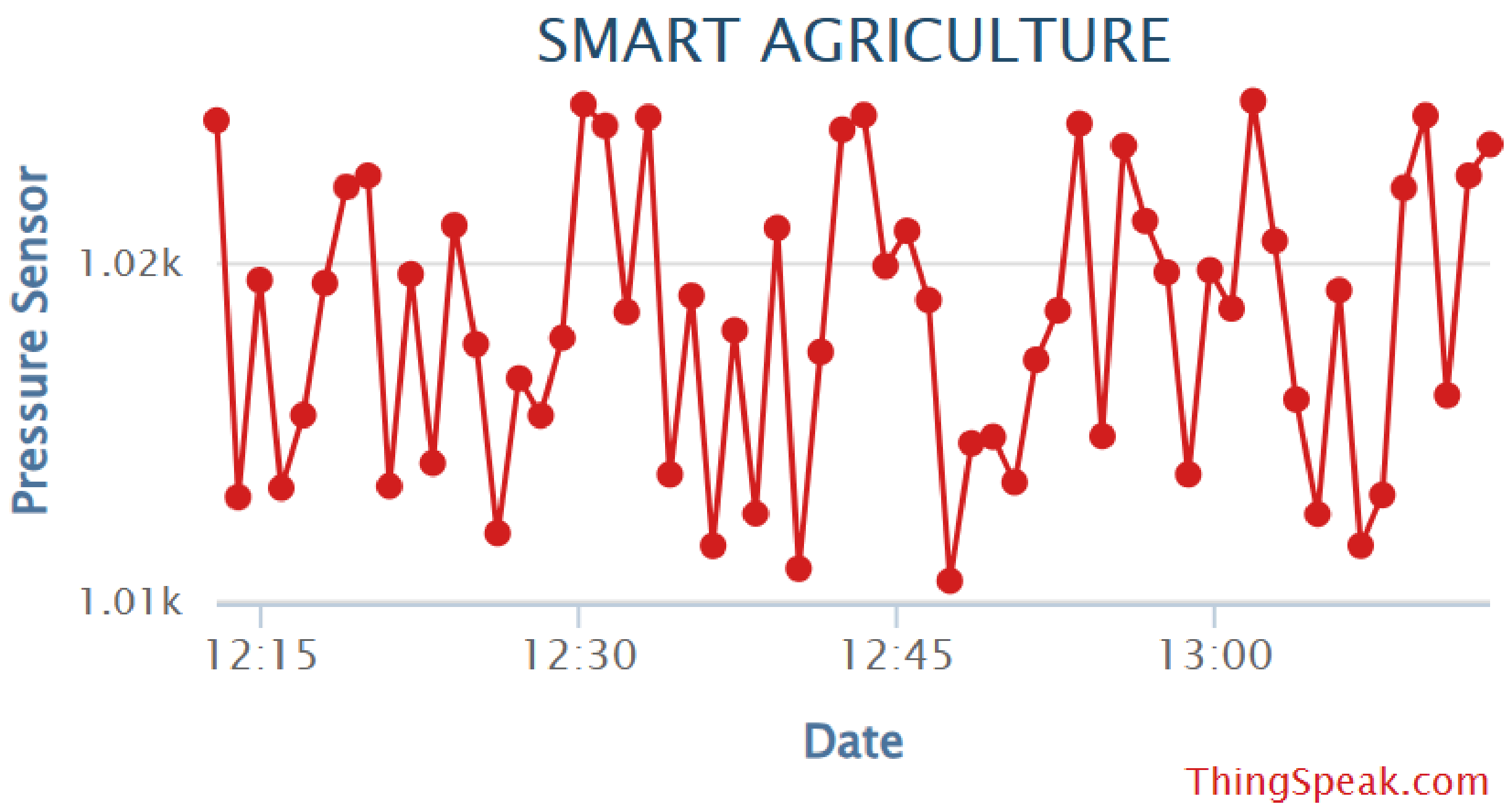

*Figure 5-2Simulated Pressure Variation Over 1 Hour*

In **Figure 5-2**, the graph of pressure variations over time reflects dynamic atmospheric conditions. The pressure values range between a minimum of approximately 1010.63 hPa and a maximum of 1024.81 hPa, with an average of 1018.14 hPa. The fluctuations indicate alternating phases of high and low-pressure systems.

For instance, a sharp increase in pressure around the third minute suggests the arrival of a high-pressure system, which could simulate clear skies and stable weather conditions. Conversely, the drop in pressure observed around the fourth minute could simulate a low-pressure system, often associated with cloudy weather or rainfall. Such variations highlight the importance of monitoring atmospheric pressure for predicting weather anomalies like storms or calm periods, which directly influence agricultural planning and crop health. These insights align with the research's focus on leveraging IoT systems for real-time weather monitoring and tailored agricultural applications.

### 5.2.3 SOIL MOISTURE DATA

Figure 4-3 shows the simulated soil moisture levels.

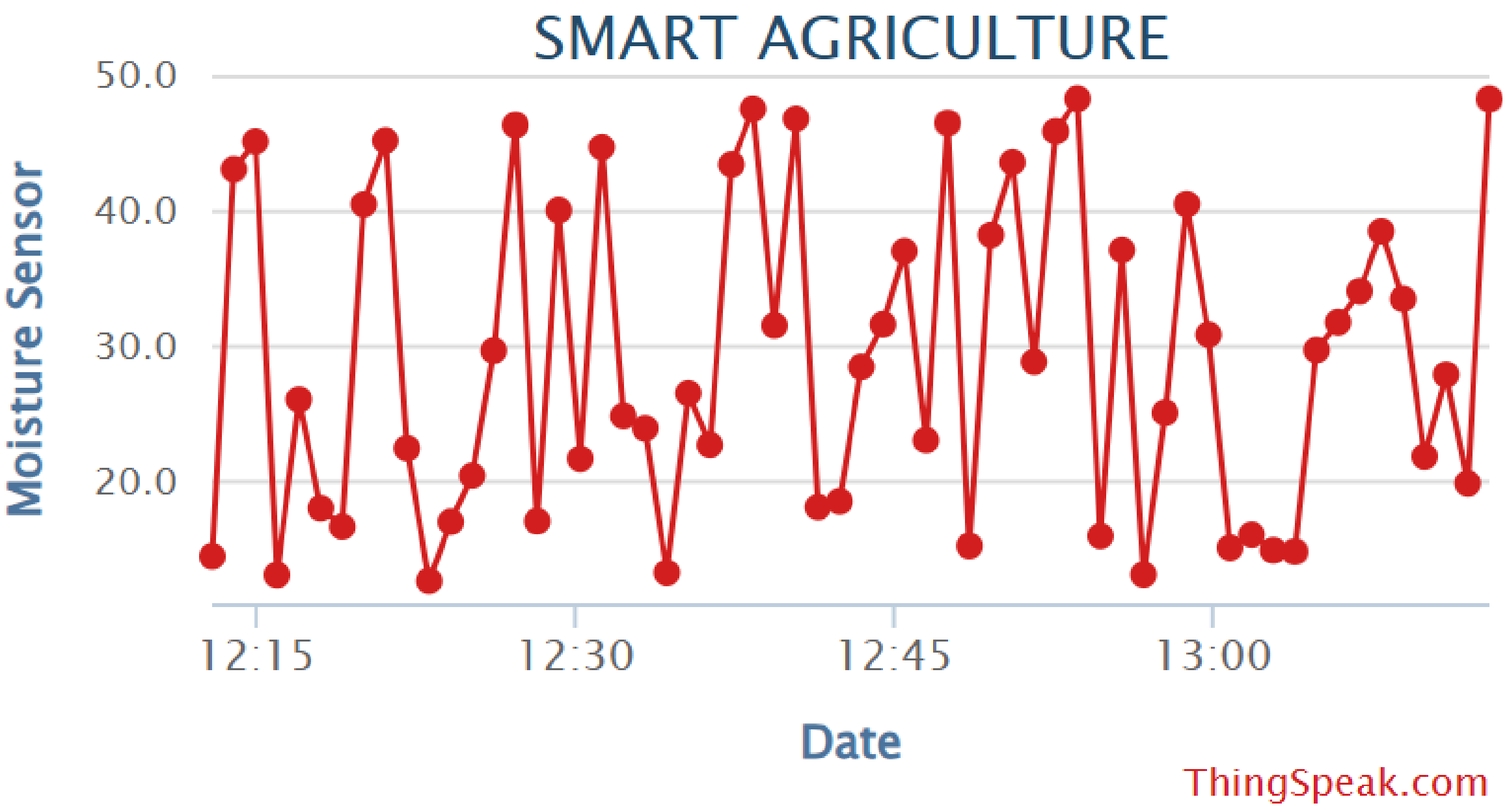

*Figure 5-3Simulated Soil Moisture Variation in 1 Hour*

In **Figure 5-3**, the graph of soil moisture variations over time illustrates the dynamic fluctuations in soil moisture levels. The moisture values range from a minimum of approximately 12.67% to a maximum of 48.34%, with an average of 29.24%. These variations reflect the shifting moisture conditions in the soil, likely influenced by environmental factors like rainfall, irrigation, and evapotranspiration.

For example, a notable increase in moisture around the 8th minute signifies a potential irrigation event or a period of rainfall, which could improve soil conditions for plant growth. On the other hand, the sharp drop in moisture observed around the 6th minute could indicate a period of soil drying, possibly due to increased evaporation or a lack of rainfall.

These fluctuations underscore the importance of real-time soil moisture monitoring, especially for agriculture, where precise moisture levels are critical for optimal crop health and yield. The insights gained from this data can be used for more efficient irrigation management, ensuring crops receive adequate water while minimizing waste.

### 5.2.4 RAINFALL DETECTION

The rain detection data provides a straightforward indication of whether rain was detected at specific times. The values represent either the presence or absence of rain, with readings that could be interpreted as "1" for rain detected and "0" for no rain.

For example, a reading of "1" suggests that rain was detected at that particular time, while a "0" indicates dry conditions. These values are crucial for smart agriculture applications, where real-time rain detection can influence automated irrigation decisions. When rain is detected (1), irrigation systems can be paused to prevent overwatering, conserving water resources. Conversely, when no rain is detected (0), irrigation systems can be activated to ensure that crops receive the necessary water for healthy growth.

## 5.3 MOBILE APP INTEGRATION

The mobile app integrates with the IoT system to display real-time weather data and send notifications based on certain environmental thresholds. This integration allows farmers to receive actionable alerts on temperature, soil moisture, pressure fluctuations, and rainfall detection.

### 5.3.1 Mobile App Interface

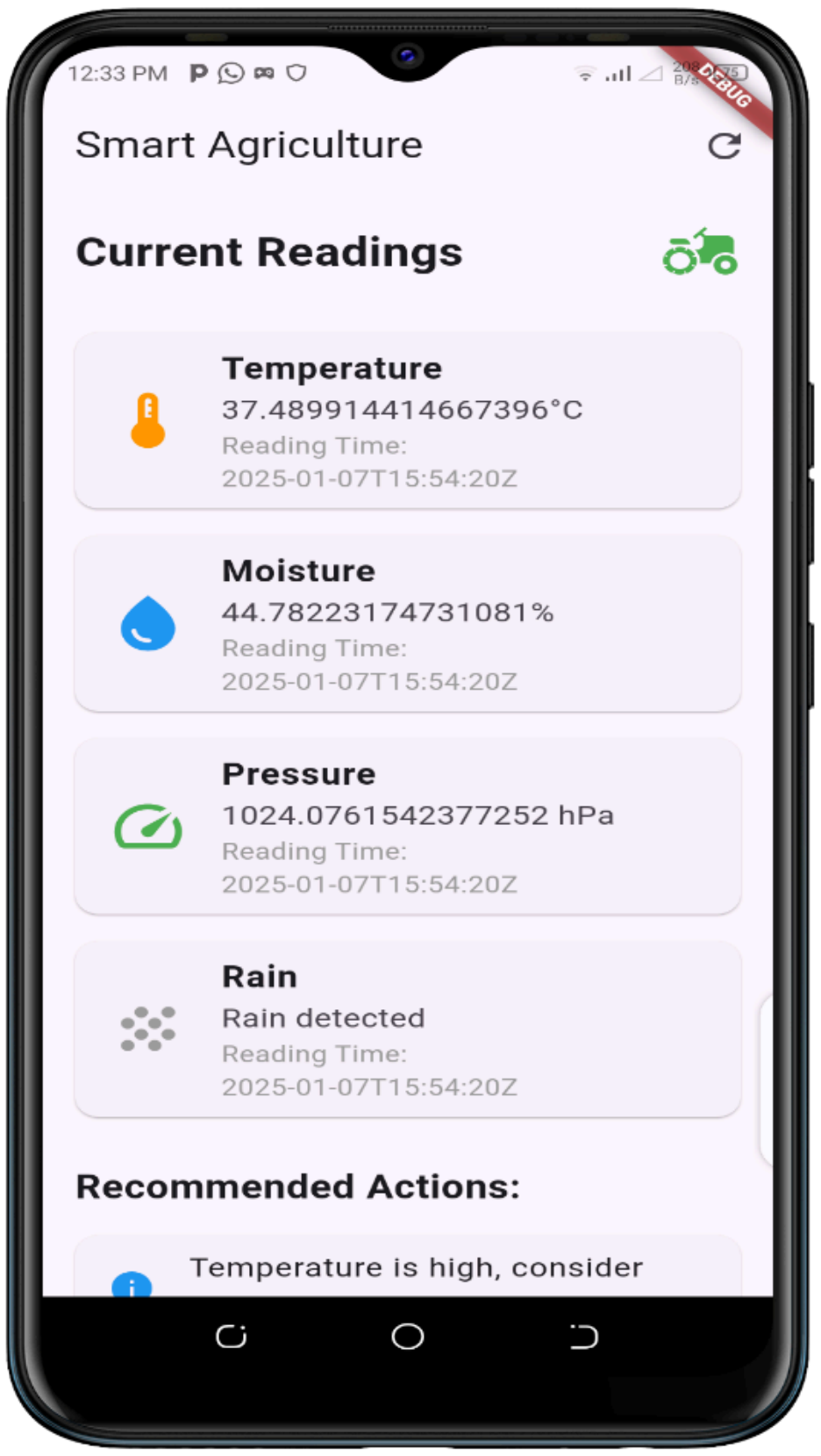

*Figure 5-4Mobile App Displaying Simulated Data*

Figure 5-4 illustrates the app's dashboard, which includes several key features to aid farmers in monitoring their environmental conditions. The dashboard displays the current temperature and pressure, using color-coded indicators for easy and immediate comprehension. It also shows real-time soil moisture levels, with an alert system that notifies the farmer when the moisture drops too low, indicating the need for irrigation. Additionally, the dashboard features rainfall detection, alerting the farmer when significant rainfall is detected and providing a trend of previous rainfall events to assist with planning and decision-making.

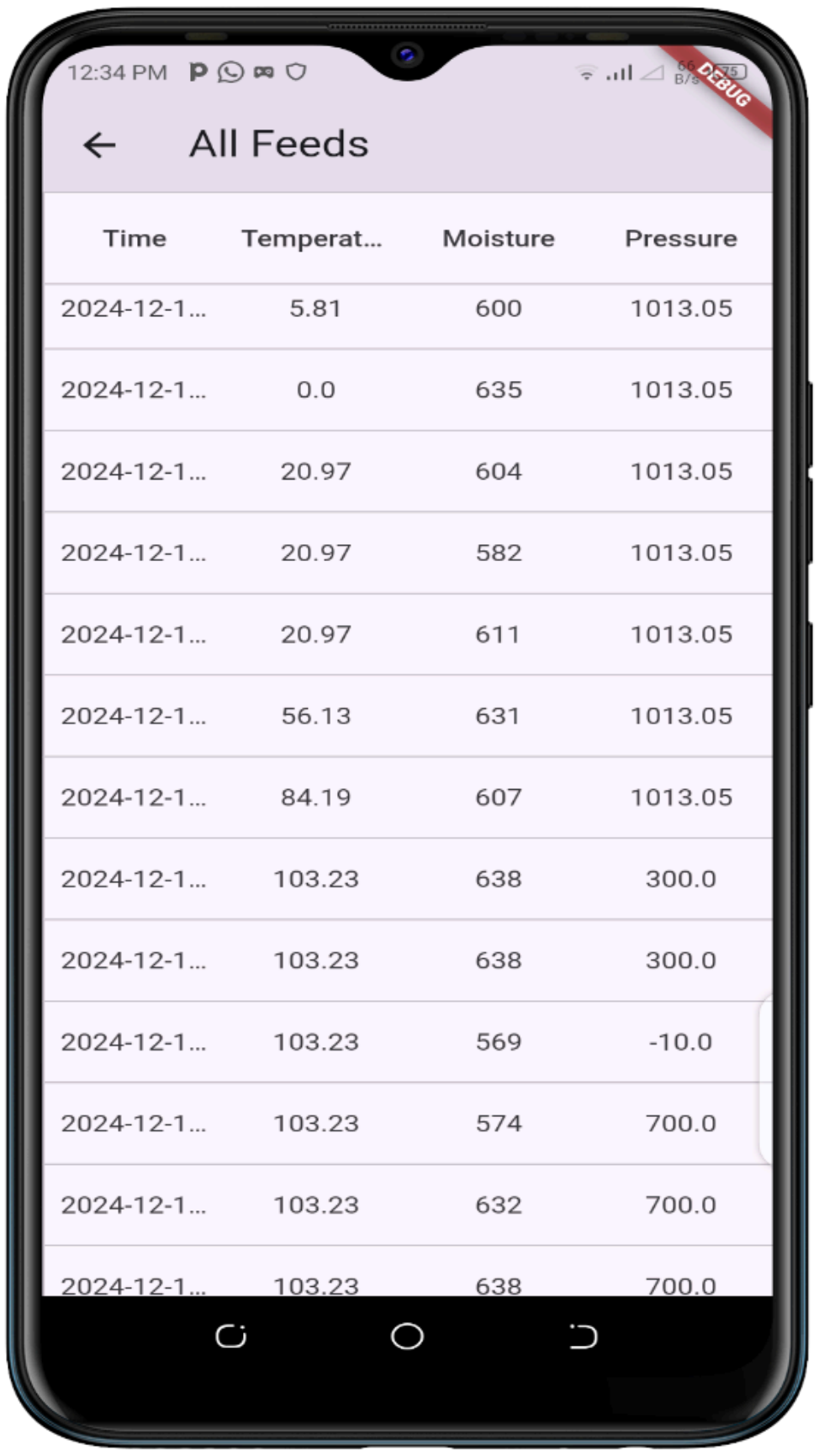

*Figure 5-5 Mobile App Displaying Simulated Data*

Figure 5-5 depicts a screen displaying historical data in a tabular format, with rows organized by date and columns representing various environmental parameters such as temperature, pressure, soil moisture, and rainfall. This data is fetched from the ThingSpeak cloud platform, ensuring that the information is up-to-date and stored securely for long-term access. By reviewing this table, farmers can gain insights into the historical trends of environmental conditions over a given period. The screen's relevance lies in its ability to provide a clear record of past data, helping farmers track changes, identify patterns, and make informed decisions about irrigation, crop management, and other agricultural practices. Accessing this data from the ThingSpeak cloud platform enhances the system's reliability and scalability, allowing farmers to analyze data remotely and optimize their farming strategies based on historical environmental trends.

### 5.3.2 Mobile App Alerts

The mobile app was programmed to trigger alerts based on specific thresholds for temperature, pressure, and soil moisture. For example, when the temperature exceeded 35°C for an extended period (on days 4 and 5), a "Temperature is high, consider cooling measures" was sent.

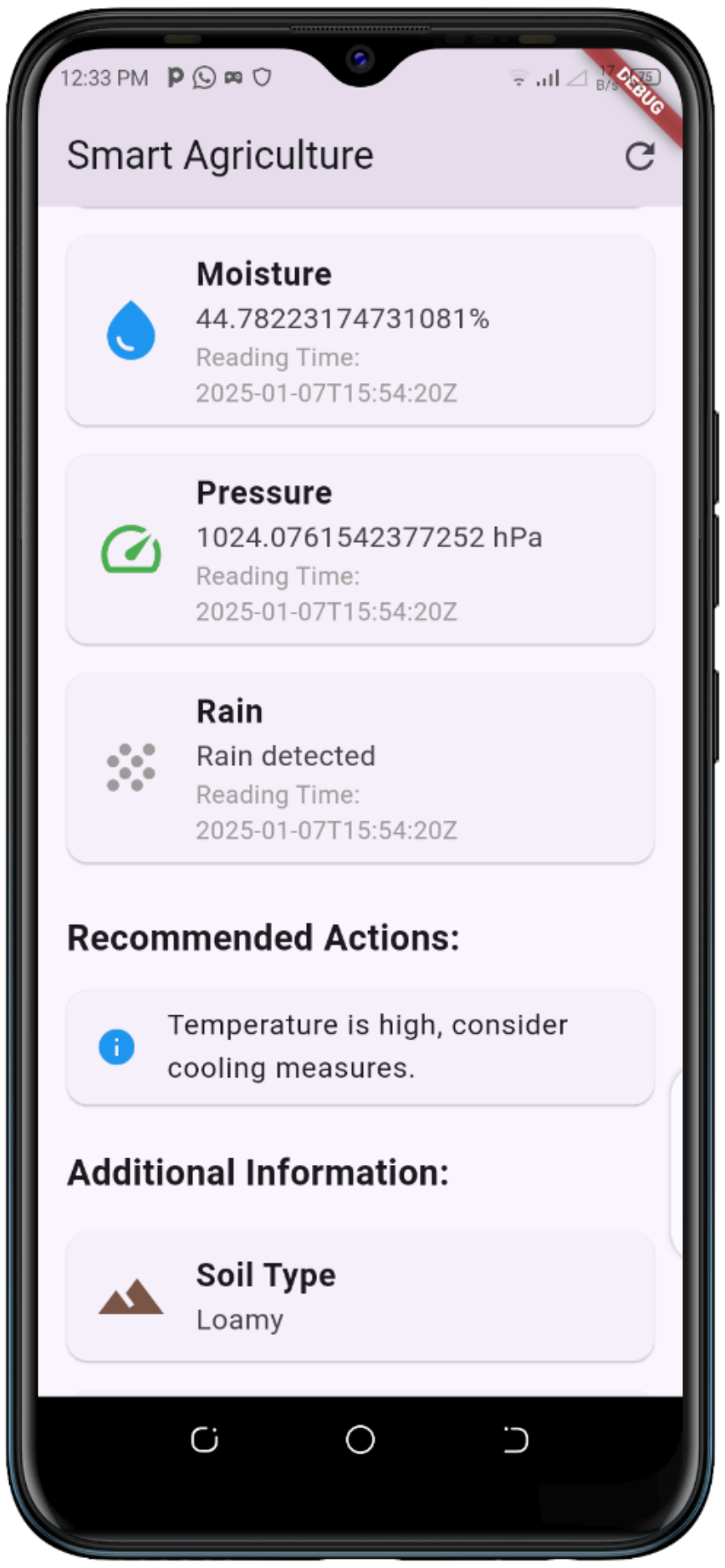

*Figure 5-6 Mobile application notification*

From figure 5-6, the key observations from the system indicate that the mobile app successfully delivered timely notifications based on the simulated environmental data, such as "Temperature is high, consider cooling measures" and "Temperature is high, consider cooling measures." These alerts played a crucial role in ensuring that the farmer took necessary actions to manage crop health and optimize water usage. The effective delivery of these notifications demonstrates the system's practical application in helping farmers make informed decisions and efficiently manage their agricultural practices, thereby improving overall productivity and resource utilization.

### 5.3.3 Forecasting and sensor control

The system also incorporates a basic forecasting mechanism using the moving average technique to provide actionable insights into weather trends, with the forecast graph displayed in a Flutter mobile application(Figure 5-7) for easy accessibility. By processing real-time data collected from sensors, the system calculates a moving average of temperature readings over a fixed window of recent data points, effectively smoothing out short-term fluctuations and highlighting broader trends. This technique is implemented by iterating through the data and averaging the temperature values within a sliding window of three consecutive data points. The forecasted values are then visualized alongside the actual data in an interactive graph within the mobile app, built using the Syncfusion Flutter Charts package. This user-friendly interface allows farmers to intuitively monitor current conditions and short-term projections, enabling them to anticipate temperature variations, optimize irrigation schedules, and make informed decisions to mitigate the adverse effects of unpredictable weather patterns.

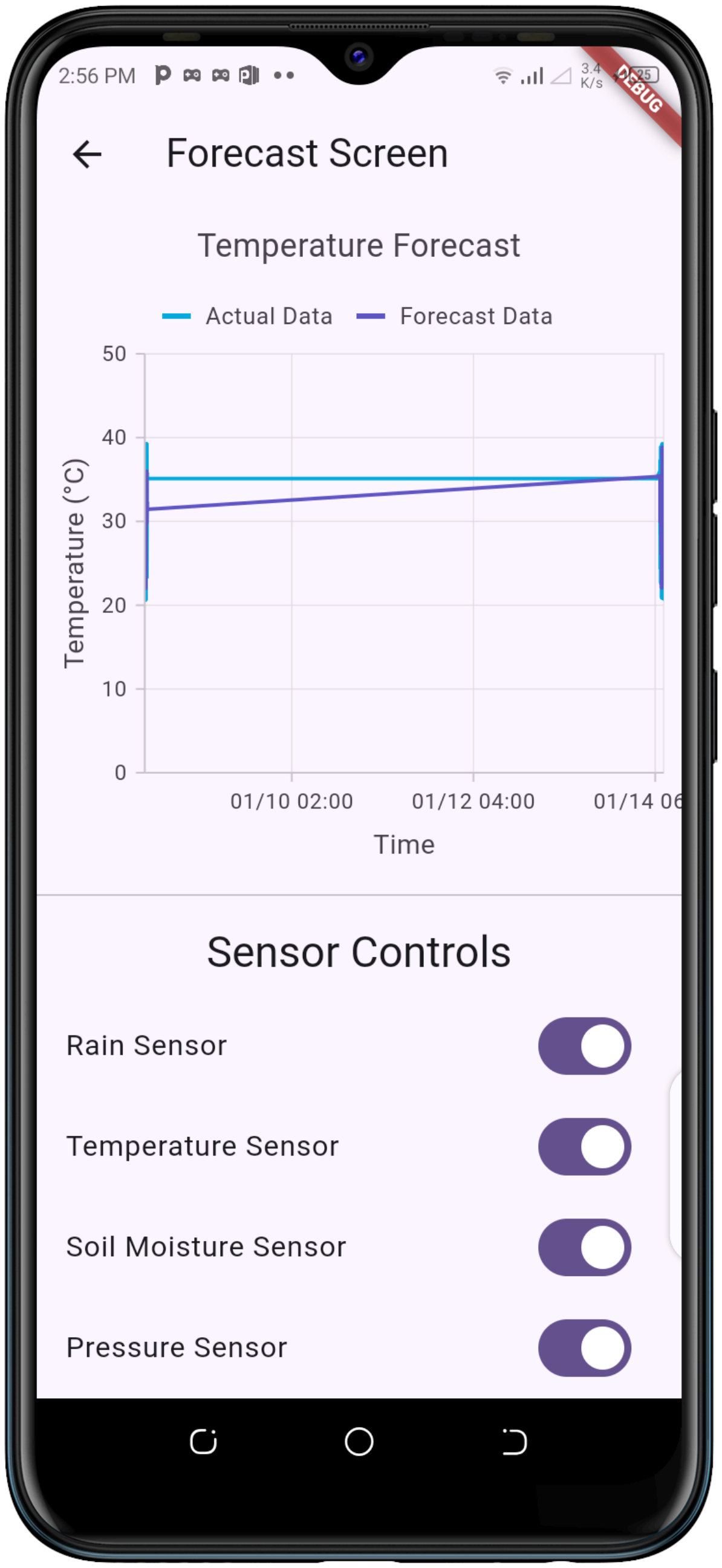

*Figure 5-7 Forecasting and sensor status*

### 5.4 SYSTEM PERFORMANCE AND INSIGHTS

The simulation results offer valuable insights into the IoT system's effectiveness in monitoring and responding to environmental conditions. Key observations highlight the system's ability to detect critical weather parameters and send timely alerts, showcasing its potential to assist farmers in making proactive decisions. The system accurately monitors pressure changes, which can signal approaching storms or weather events, while temperature monitoring ensures farmers are alerted to heat stress conditions that could negatively affect crop yield. Additionally, the integration of rainfall detection and soil moisture management allows the system to provide real-time irrigation recommendations, optimizing water usage and helping farmers maintain healthy crops.

### 5.5 LIMITATIONS AND FUTURE ENHANCEMENTS

While the simulation provided valuable insights, there are some limitations to consider. The simulation is based on predefined data, meaning that real-world systems will encounter more variability, including sensor noise or unexpected weather events. Additionally, the simulated system does not account for sensor calibration or potential errors that could arise with real hardware. Future improvements will focus on enhancing the system’s accuracy by incorporating real-time weather data, expanding sensor coverage, and refining the mobile app’s functionality to provide more personalized recommendations based on user feedback.

## 6. DISCUSSIONS AND REFLECTIONS

Traditional data collection methods only displayed data on an LCD or desktop, but this research demonstrates the benefits of real-time data collection using a mobile application, offering flexibility in crop monitoring and management [17]. Farmers can access critical information remotely, enabling proactive decision-making based on changing environmental conditions. By monitoring temperature, pressure, soil moisture, and rain sensing, the IoT system ensures efficient irrigation, reduces water waste, and optimizes resource use, leading to increased yields and sustainability [18]. Integrating forecasting further enhances this system by predicting weather patterns and allowing sensors to be turned off during periods of inactivity, such as when no rain is expected. This reduces energy consumption, extends sensor lifespan, and prevents unnecessary irrigation [19]. However, this approach diverges from some previous research, which mainly focuses on continuous monitoring without incorporating predictive forecasting to manage sensor activity. Traditional IoT systems often operate under the assumption that sensors must always be on for real-time data collection, disregarding energy and resource optimization. The difference here lies in the dynamic management of sensors based on forecasted conditions, offering both cost savings and greater sustainability [20].A key limitation of this research is that it was conducted in an experimental setup, not using real-world devices, which may pose scalability and calibration challenges in diverse environments. Additionally, the basic forecasting model used may not capture weather patterns with high accuracy. Future work will focus on real-world implementation, addressing scalability and calibration issues, and enhancing the forecasting system with advanced machine learning models. We will also investigate the long-term impact of the IoT system on farm profitability, exploring how efficient irrigation and resource optimization translate into economic benefits for farmers [21].